\documentclass[a4paper]{article}
\usepackage{graphicx}
\usepackage{subcaption}
\usepackage{INTERSPEECH2022}
\newcommand{\argmax}[1]{\underset{#1}{\operatorname{arg}\,\operatorname{max}}\;}

\title{Audience Response Prediction from Textual Context}
\name{Ibrahim Shoer, Berker Türker, Engin Erzin}
\address{
  KUIS AI Lab., College of Engineering, Ko\c{c} University, Istanbul, Turkey}
\email{[ishoer20,bturker13,eerzin]@ku.edu.tr}
\begin{document}

\maketitle
\begin{abstract}
Humans' perception system closely monitors audio-visual cues during multiparty interactions to react timely and naturally. Learning to predict timing and type of reaction responses during human-human interactions may help us to enrich human-computer interaction applications. In this paper we consider a presenter-audience setting and define an audience response prediction task from the presenter's textual speech. The task is formulated as a binary classification problem as occurrence and absence of response after the presenter's textual speech. We use the \emph{BERT} model as our classifier and investigate models with different textual contexts under causal and non-causal prediction settings. While the non-causal textual context, one sentence preceding and one sentence following the response event, can hugely improve the accuracy of predictions, we showed that longer textual contexts with causal settings attain UAR and $F1$-Score improvements matching and exceeding the non-causal textual context performance within the experimental evaluations on the OPUS and TED datasets.
\end{abstract}

\section{Introduction}
\label{sec:intro}

Audience responses, such as cheers, claps, laughter or bodily gestures, generate interactivity with the presenter and define a key asset for the assessment of interaction quality. Learning an understanding for the timing and type of audience responses may help to assess interaction quality as well as to synthesize timely and proper response generation. For example, audience responses are inserted in sitcoms to grab the audience's attention, making them follow up and react accordingly. Deep learning models can be deployed to precisely predict audience responses over large datasets, which can be constructed from caption data such as the Open Subtitles Dataset (OPUS) \cite{lison-tiedemann-2016-opensubtitles2016}. 

Audience responses mainly rely on the context, which can be defined as the uttered text preceding the response as well as the following text. Several deep learning models were presented for laughter prediction in sitcoms using words, word2vec and character trigrams \cite{bertero2016long}, using audio and language features \cite{7472785}, and using audio, video and text modalities \cite{patro2021multimodal}. These models consider the preceding utterance as the context to define a causal laughter prediction model. On the other hand, using the following utterance in the context can leverage the classification performance, and define a practical upper bound for the prediction task. A third factor would be the length of the context history. We should expect improvements in prediction performance as the context history gets longer with a likely performance saturation as context gets uncorrelated with the response event.

In this study we set our goal as to predict audience response from the presenter's textual speech. We formulate the task as a binary classification problem; whether a response occurs (+) or not (-) right after the presenter's textual speech. In order to train and test the binary classifier, we build a balanced dataset of (+) and (-) response event classes together with the associated textual context from TV shows, where (+) response events are set as laughter, sobbing and cheering instances. Lately, language models trained for multiple tasks and on large datasets can outperform classical machine learning algorithms on many tasks \cite{gonzalez2020comparing}. \emph{BERT} is a pre-trained model for the next sentence prediction that can be fine-tuned for context labeling tasks. We use the \emph{BERT} model as our classifier and investigate roles of different textual context settings for the binary classification task. Textual context unit is set as a sentence and the context is organized as a sequence of sentences. Different length textual contexts under causal and non-causal settings are investigated for the audience response prediction.
Main contributions of this study are listed as:
\begin{itemize}
\setlength\itemsep{0em}
 \item We propose a \emph{BERT} based audience response prediction model.
 \item We define a standard model with single sentence textual context preceding the response event.
 \item We show that using the non-causal textual context, one sentence preceding and one sentence following the response event, can hugely improve the accuracy of predictions.
 \item We investigate the textual context length with the causal setting and observe performance improvements matching and exceeding the non-causal textual context performance. 
\end{itemize}




\section{Literature Review}
\label{sec:lit}
Audience response is characterized by many factors, whereas timing and the type of the response are two important factors. Several studies have been conducted on the relation between applause and rhetorical devices \cite{atkinson_1985} \cite{heritage1986generating}, where rhetorical devices are defined by actions done by the presenter including: contrasts, puzzle-solution,  headline-punchline and  combination. Certain parts of speech can trigger similar responses like greeting phrases usually trigger clapping \cite{bull2011invitations,liu2017fostering}. 

Audience response is also a subject of sitcoms where the responses are placed at certain moments of the episodes. Responses are usually laughter effects or pre-recorded laughter sounds whereas other responses (clapping, cheering etc.) are rarely used. Few studies focuses on predicting audience laughter timings in TV sitcom episodes. In \cite{bertero2016deep}, Bertero and Fung used audio and language features for humor prediction over the dialogues of "The Big Bang Theory" TV show. They compared performances of CRF, CNN, RNN deep learning structures and reported CNN as the best performer with 68.5\% F-Score. 
Also, they reported performance improvements by using LSTM in \cite{bertero2016long} and by multimodal approach in \cite{bertero2016multimodal}. In another study \cite{patro2021multimodal}, Patro et al. made an effort to the similar problem and argued that existing LSTM and BERT based networks with  only language features do not perform as well as joint text and video or only video-based solutions. They used text, audio, and video modalities for predicting laughter tracks on the aforementioned TV show. They reported 79.96\%, 79.30\%, 81.32\% F-Scores for text only, video only and fusion respectively. Kayatani et al., in their study \cite{kayatani2021laughing}, prepared a new dataset for the show and they include timestamps of laughter instances over the episodes. The study focuses on textual, facial expression, and character modalities, claiming that knowing the speaking character improves prediction performance.

Choube and Soleymani \cite{choube2020punchline} suggested context-aware hierarchical multimodal fusion with visual, textual and acoustic channels. They experiment with baselines and their methodology over the UR-FUNNY dataset and reported the highest F-Score as 68.85\%. Hasan et al. proposed Humor Knowledge enriched Transformer (HTK) \cite{hasan2021humor}. They incorporate humor centric external knowledge besides visual and acoustic channels by employing separate transformer encoders. They reported 77.36\% and 79.41\% accuracy in humorous punchline detection over UR-FUNNY and MUStaRD datasets.

On the other hand, there are several works using single modality which is text, usually short texts, in order to detect humor. Fan et al. suggested internal and external attention neural network \cite{fan2020humor} for the humor detection in which two types of attention mechanisms are integrated to capture the incongruity and ambiguity in humor text. In their other work \cite{fan2020phonetics}, they proposed Ambiguity Comprehension Gated Attention network to learn phonetic structures and semantic representation for humor recognition. 
They showed the benefits of their contributions over two public short text datasets.


Navarrete et al.  proposed an audience response prediction system in a political setting from Barack Obama's speech, based on speech, silent pauses and co-gestures \cite{navarretta2017prediction}. They indicated that Obama uses pauses especially to emphasize his jokes and let the audience get the point. Similarly, Ruf and Navarrette focuses on gestures (head movements, hand gestures etc.) and obtain experimental results over Donald Trump's speech sessions \cite{ruf2020creating}. 

In the perspective of audience response, the real life scenarios and instances such as political speech in front of a crowd, TED Talks are valuable data sources for machine learning. Liu et al.  extracted the "applause" instances from the TED talks dataset, studied the applause generation and proposed hypothesized rhetorical devices for applause generation \cite{liu2017fostering}.

In this work, we use only textual features with BERT \cite{devlin2018bert} which is a generalized language model pre-trained on a huge dataset. In the training of downstream task, audience response prediction, we take advantage of an extensively large dataset sourced from OPUS \cite{lison-tiedemann-2016-opensubtitles2016} and TED Talks \cite{yoon2019robots}. Causal and non-casual prediction schemes are investigated on both datasets. Instead of only humor or laughter prediction, we deal with various kinds of additional audience responses such as sigh, cry, scream (see Table~\ref{tab:dataset_dist}). 


\section{Methodology}
\label{sec:method}
Modeling and predicting audience response is a valuable skill for an engaging agent in human-computer interaction scenarios. In this study, we propose an audience response prediction model from the presenter's textual speech. While our model defines a binary classification task to predict response occurrence (+) and (-) non-occurrence, we test the performance of the model with different length textual contexts under causal and non-causal settings.

\subsection{Approach}
\label{ssec:approach}
We take the textual context unit as a sentence and the context is organized as a sequence of sentences. Hence, the sequence of sentences for the presenter can be defined as
\begin{equation}
\mathcal{S} = \{...,s_{t-1},s_{t},s_{t+1},...\}
\end{equation}
where $s_t$ is the sentence uttered at present time $t$. Then, an audience response event $e_t$ can be defined to occur or not to occur preceding $s_t$ and following $s_{t+1}$ as\begin{equation}
e_t =\begin{cases} + & \text{response event} \\ - & \text{no response} \end{cases}
\end{equation}

Then the audience response prediction problem can be defined as
\begin{equation}
\hat{e}_t = \argmax{e} \text{ARP}^m_n(e|s_{t-n},s_{t-n+1},......,s_{t+m}) 
\end{equation}
where $\text{ARP}^m_n$ is the audience response prediction network with textual context from sentence $s_{t-n}$ to $s_{t+m}$.

A set of different context configurations are used in the experimental evaluations. We set single sentence response prediction model ($\text{ARP}_0^0$) as the standard model, where the task to predict response occurrence $\hat{e}_t$ by observing the sentence $s_t$ at time $t$. Furthermore, knowing the future is expected to bring important evidence for the occurrence of the response. Hence, we define a non-causal target predictor as $\text{ARP}_0^1$, which performs prediction for $\hat{e}_t$ by observing the sentence $s_t$ and $s_{t+1}$. Although such a non-causal predictor is often not useful for timely response prediction, it can define a performance target for the causal predictors ($\text{ARP}_n^0$). In the experimental evaluations we test textual context up to sentence lengths of $n=4$.

\subsection{Model}
\label{ssec:model}
While transformer-based language models are trained for text completion, masked language and next sentence prediction, they can also be fine-tuned for various tasks. Textual modality from the spoken sentences can be modeled using these language models. The self-attention technique in transformer models combines information from all elements of a sequence into context-aware representations. These representations can infer a likelihood for the audience response. In this study, we use the \emph{BERT} \cite{devlin2018bert} model for this inference task.

\begin{figure}[ht]
 \centering
 \includegraphics[scale=0.24]{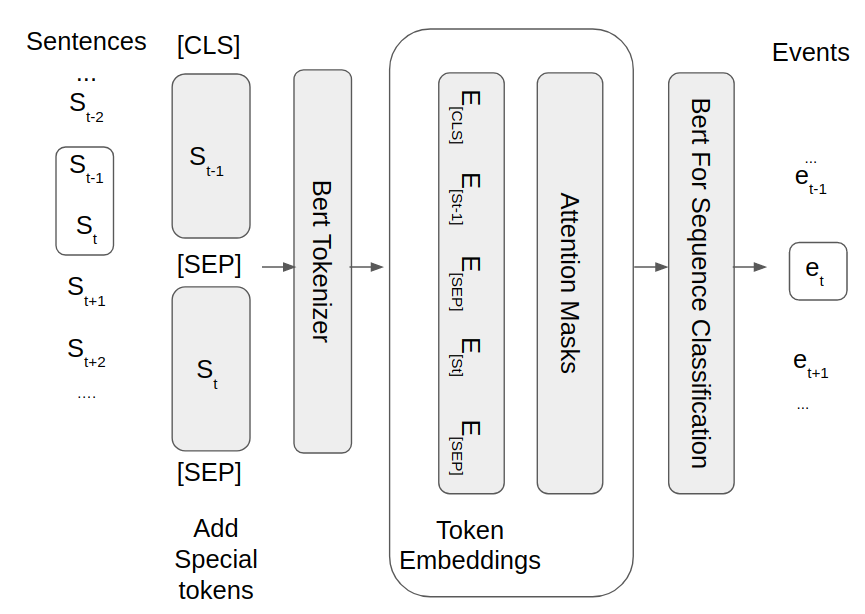}
 \caption{Audience response prediction model}
 \label{fig:workflow}
\end{figure}
For this task, we fine-tune a 12 layer encoder transformer based \emph{BERT} model, where each encoder layer contains a self-attention layer, add and normalize layer and a feed forward layer. The 12-layer BERT model is used with an uncased vocab from the transformers library, with two labels at the output. In \emph{BERT}, the first token of the last layer, the \emph{CLS} token is reserved for classification. At the input, \emph{SEP} token is used to indicate the end of each sentence. \emph{BERT} uses word-piece tokenizer, breaking each word into sub-words. After tokenization, we encode the tokens into word embedding and attention masks.

For training, we use Adam optimizer with learning rate of 2e-5, epsilon as 1e-8 and all models are fine-tuned for four epochs. All experiments were done on Tesla V100 GPU with memory 32GB. Figure~\ref{fig:workflow} illustrates the model. In the model configuration, sentences are associated using the [SEP] token, the [CLS] token is appended at the start and [SEP] token at the end; in the figure, two preceding sentences $S_{t-1}$ and $S_t$ are associated together with the special tokens  to predict the likelihood of the (+) response event occurrence for $e_t$. Bert tokenizer is then used to retrieve the sentences token embeddings (indices of input sequence tokens in the vocabulary) and attention masks (mask to avoid performing attention on padding token indices), both embeddings and attention masks are fed to \emph{BERT} model, the first token of the output embedding indicates the (+) response event occurrence. 

\section{Experimental Evaluations}
\label{experimental}

\subsection{Datasets}
\label{ssec:Datasets}
We set audience responses as diverse as possible including laughter, sobbing and cheering instances. A large set of response events are used from the Open Subtitles Dataset (OPUS) \cite{lison-tiedemann-2016-opensubtitles2016}. Each line in the dataset is considered as a uttered sentence. For the lines containing an event in the middle, words from the beginning of the line till the event are considered the preceding turn, and words after the event till the end of the line are considered the following turn. We make sure that no response or any hearing impaired notation is present in the textual context.

\begin{table}[h]\centering
\caption{Distribution of (+) response events in the ARC from the OPUS dataset}\label{tab:dataset_dist}
\renewcommand{\arraystretch}{1.2}
\begin{tabular}{ c  c }
\toprule
Responses & \# of occurrences \\ \hline\hline
clap & 8811 \\ \hline
applause &17455 \\ \hline
cheer & 76377 \\ \hline
chuckle & 237996 \\ \hline
cry & 39010 \\ \hline
laugh & 362991 \\ \hline
scream & 161390 \\ \hline
shout & 76980 \\ \hline
sigh & 240409 \\ \hline
grunt & 204236 \\ \hline 
sob & 45486 \\ \hline\hline
Total: & 1471141 \\ \bottomrule
\end{tabular}
\end{table}

To evaluate the audience response prediction in presenter-audience setting, we utilize the TED gesture dataset, which is a large-scale English-language dataset of TED talks videos \cite{yoon2019robots}. We collected the transcriptions containing 5975 (+) response and 55642 (-) response events. Models fine-tuned on OPUS dataset are then fine-tuned on the TED dataset for one epoch.

\subsection{Evaluation metrics}
\label{ssec:evaluation}
Since the response prediction task is defined as a binary classification problem, the receiver operating characteristic curve (ROC) would be a valuable objective evaluation metric. Furthermore, we also report unweighted average recall (UAR), $F1$-Score, and the area under the ROC curve (AUC) metrics.

The UAR metric is defined as the average of (+) response and (-) response recall rates, as
\begin{equation}
\text{UAR} = (R^+ + R^-)/2.
\end{equation}
The recall rates $R^+$ and $R^-$ are defined as
\begin{equation}
R^+ = \frac{TP}{TP + FN}, ~~~~ 
R^- = \frac{TN}{TN + FP}
\end{equation}
where $TP$ is true positive, $FN$ is false negative, $TN$ is true negative and $FP$ is false positive. Recall that $F1$-Score is defined as the harmonic mean of recall and precision and calculated as
\begin{equation}
F1 = \frac{TP}{TP + (FP + FN)/2}.
\end{equation}
Hence monitoring UAR and $F1$-Score has indications for the precision as well.

\subsection{Results}
\label{sec:Results}
Experimental evaluations are carried over the balanced OPUS and unbalanced TED datasets. The single sentence response prediction model $\text{ARP}_0^0$ is set as the standard and the non-causal response prediction model $\text{ARP}_0^1$ is set as the target. Knowing the future should bring important evidence for the occurrence of the response. Hence the non-causal model can define a performance target for the causal predictors ($\text{ARP}_n^0$). We test the textual context with the causal predictors up to sentence lengths of $n=4$. Furthermore, we incorporate the Naive Bayes algorithm, which is used in \cite{navarretta2017prediction} by Navarrete et al. for audience response prediction, as a baseline.


\begin{figure*}[ht]
\centering
\begin{subfigure}{.5\textwidth}
  \centering
  \includegraphics[scale=0.5]{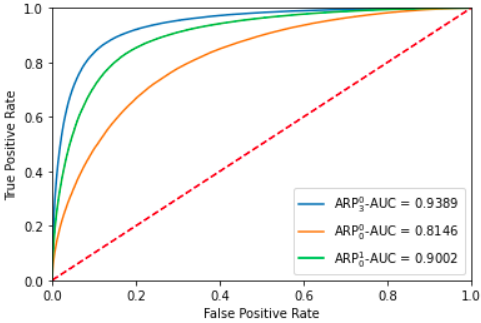}
  \caption{ROC curves over the balanced OPUS dataset}
  \label{fig:sub1}
\end{subfigure}%
\begin{subfigure}{.5\textwidth}
  \centering
  \includegraphics[scale=0.48]{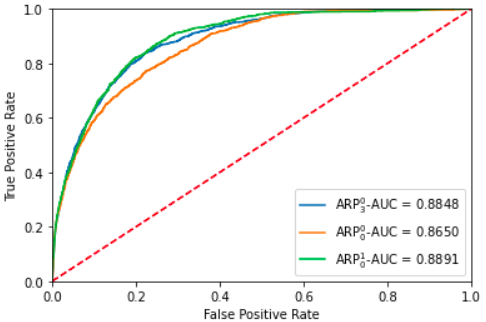}
  \caption{ROC curves over the unbalanced TED dataset}
  \label{fig:sub2}
\end{subfigure}
\caption{Receiver Operating Characteristic (ROC) curves for the standard $\text{ARP}_0^0$, the non-causal target $\text{ARP}_0^1$, and the causal predictor with context length of 4, $\text{ARP}_3^0$, over the OPUS and TED datasets}
\label{fig:test}
\end{figure*}

Figure~\ref{fig:sub1} presents the  Receiver Operating Characteristic (ROC) curves for the standard $\text{ARP}_0^0$, the non-causal target $\text{ARP}_0^1$, and the causal predictor with context length of~4, $\text{ARP}_3^0$, over the balanced OPUS dataset. While the target $\text{ARP}_0^1$ performs significantly better compared to the standard $\text{ARP}_0^0$, the causal predictor with context length of~4 outperforms the target predictor clearly. This is an encouraging observation that longer causal context can predict audience response better than the non-causal target.

Table~\ref{tab:opus-results} presents UAR, AUC and $F1$-Score performances of the causal response predictors $\text{ARP}_n^0$ compared to the standard and the target over the OPUS dataset. The UAR performances of the Naive Bayes baseline, $UAR_{NB}$, are also given in Table~\ref{tab:opus-results}.
It is clear that performance improvements are consistent as $n$ gets larger. Starting from $n=2$, $\text{ARP}_n^0$ performs better than the non-causal target $\text{ARP}_0^1$. Although incremental performance improvements getting smaller as $n$ gets larger, we still observe more than 1\% improvements for all metrics with $n=4$. This is likely due to the rich and large ARC collected from the OPUS dataset that provides better training performance with increasing context length. Although the Naive Bayes baseline performance is also improving as context gets larger, it sustains a significantly lower UAR performance compared to the proposed ARP models.
\begin{table}[ht]\centering
\caption{Response prediction performance of the proposed networks together with the Naive Bayes baseline UAR on the balanced ARC from the OPUS dataset}
\label{tab:opus-results}
\renewcommand{\arraystretch}{1.2}
\begin{tabular}{ c  c  c  c  c }\hline
Network & $UAR_{NB}$ & $UAR$ & $AUC$ & $F1$-Score\\  \hline\hline
{$\text{ARP}_0^0$ }& 62.76 & 74.00 & 81.46 & 74.43 \\ \hline
{$\text{ARP}_0^1$ }& 67.51 & 82.80 & 90.02 & 82.93 \\ \hline
{$\text{ARP}_1^0$ }& 66.28 & 80.66 & 88.24 & 80.93\\ \hline
{$\text{ARP}_2^0$ }& 68.75 & 84.89 & 91.96 & 85.01\\ \hline
{$\text{ARP}_3^0$ }& 70.71 & 87.14 & 93.89 & 87.23 \\ \hline 
{$\text{ARP}_4^0$ }& 72.48 & 88.82 & 95.15 & 88.88 \\ \hline
\end{tabular}
\end{table}

Figure~\ref{fig:sub2} presents the  ROC curves for the standard $\text{ARP}_0^0$, the non-causal target $\text{ARP}_0^1$, and the causal predictor with context length of 4, $\text{ARP}_3^0$, over the unbalanced TED dataset. Models are first pre-trained on the OPUS, then fine-tuned over the training partition and evaluated on the test where train and test splits are respectively 80\% and 20\% over the TED dataset. Performance tendency is partially different than the ones over the OPUS dataset. The target $\text{ARP}_0^1$ performs better than $\text{ARP}_0^0$. On the other hand, performances of the causal predictor with context length of 4 and the target predictor are similar. Table~\ref{tab:ted-results} presents UAR, AUC and $F1$-Score performances over the TED dataset for balanced and unbalanced testing. Although context length 2 and greater performs better than the standard $\text{ARP}_0^0$, performance improvements are not consistent as in the OPUS dataset. Compared to the OPUS dataset, while UAR performance is holding high above 79\% for balanced and unbalanced testings, and $F1$-Score is holding high for balanced testing, especially $F1$-Score performance drop for the unbalanced testing is noticeable. This indicates lower precision performance on the unbalanced testing, which yields higher false positives for the unbalanced TED dataset. 

\begin{table}[ht]\centering
\caption{Response prediction performance of ARP \\ models fine-tuned on the unbalanced and tested\\ on the balanced/unbalanced TED dataset}
\label{tab:ted-results}
\renewcommand{\arraystretch}{1.2}
\begin{tabular}{ c  c  c  c }\hline
 Network & UAR & AUC & $F1$-Score\\  \hline\hline
{$\text{ARP}_0^0$} & 76.76 / 77.33 & 86.18 / 86.5 & 76.84 / 39.87 \\ \hline
{$\text{ARP}_0^1$} & 81.30 / 81.10 &89.15 / 88.91 & 81.24 / 44.69 \\ \hline\hline
{$\text{ARP}_1^0$ } & 79.43 / 79.50 & 87.59 / 87.61& 79.25 / 42.44\\ \hline
{$\text{ARP}_2^0$ } & 80.38 / 78.35 & 88.45 / 88.38 & 80.27 / 46.77 \\ \hline
{$\text{ARP}_3^0$ } & 80.43 / 80.51 & 88.65 / 88.48 &80.36 / 45.00\\ \hline 
{$\text{ARP}_4^0$ } & 79.93 / 80.09 & 88.22 / 88.22 &79.80 / 43.71\\ \hline
\end{tabular}
\end{table}
\section{Conclusions}
Audience response prediction is a valuable tool for the speaker that gives him confidence in the speech. In this paper, we train a language model to predict the occurrence of audience responses. Modeling textual context requires having transcriptions of the uttered speech. We investigated a set of  different context configurations on a BERT driven classification model. Results indicate that extending the length of causal textual context improves the performance much like the non-causal context. While the context in this paper is based on textual modality capturing uttered sentence sequences, other modalities, such as acoustic speech, have the potential to bring complementary information to improve audience response prediction performance. Although the balanced test scenarios result in high UAR and $F1$-Score performances, under the more likely unbalanced testing conditions we observed degradation on the $F1$-Score and precision performances. Another key observation is the consistent performance improvements for longer context lengths  on the OPUS dataset, it is not the case on the TED dataset. This suggests that the larger and richer OPUS dataset keeps training ARP models better as context lengths get longer.

\label{sec:conclusions}

\newpage
\bibliographystyle{IEEEtran}
\bibliography{template}

\begin{thebibliography}{10}
\providecommand{\url}[1]{#1}
\csname url@samestyle\endcsname
\providecommand{\newblock}{\relax}
\providecommand{\bibinfo}[2]{#2}
\providecommand{\BIBentrySTDinterwordspacing}{\spaceskip=0pt\relax}
\providecommand{\BIBentryALTinterwordstretchfactor}{4}
\providecommand{\BIBentryALTinterwordspacing}{\spaceskip=\fontdimen2\font plus
\BIBentryALTinterwordstretchfactor\fontdimen3\font minus
  \fontdimen4\font\relax}
\providecommand{\BIBforeignlanguage}[2]{{%
\expandafter\ifx\csname l@#1\endcsname\relax
\typeout{** WARNING: IEEEtran.bst: No hyphenation pattern has been}%
\typeout{** loaded for the language `#1'. Using the pattern for}%
\typeout{** the default language instead.}%
\else
\language=\csname l@#1\endcsname
\fi
#2}}
\providecommand{\BIBdecl}{\relax}
\BIBdecl

\bibitem{lison-tiedemann-2016-opensubtitles2016}
\BIBentryALTinterwordspacing
P.~Lison and J.~Tiedemann, ``{O}pen{S}ubtitles2016: Extracting large parallel
  corpora from movie and {TV} subtitles,'' in \emph{Proceedings of the Tenth
  International Conference on Language Resources and Evaluation
  ({LREC}'16)}.\hskip 1em plus 0.5em minus 0.4em\relax Portoro{\v{z}},
  Slovenia: European Language Resources Association (ELRA), May 2016, pp.
  923--929. [Online]. Available: \url{https://aclanthology.org/L16-1147}
\BIBentrySTDinterwordspacing

\bibitem{bertero2016long}
D.~Bertero and P.~Fung, ``A long short-term memory framework for predicting
  humor in dialogues,'' in \emph{Proceedings of the 2016 Conference of the
  North American Chapter of the Association for Computational Linguistics:
  Human Language Technologies}, 2016, pp. 130--135.

\bibitem{7472785}
------, ``Predicting humor response in dialogues from tv sitcoms,'' in
  \emph{2016 IEEE International Conference on Acoustics, Speech and Signal
  Processing (ICASSP)}, 2016, pp. 5780--5784.

\bibitem{patro2021multimodal}
B.~N. Patro, M.~Lunayach, D.~Srivastava, Sarvesh, H.~Singh, and V.~P.
  Namboodiri, ``Multimodal humor dataset: Predicting laughter tracks for
  sitcoms,'' in \emph{Proceedings of the IEEE/CVF Winter Conference on
  Applications of Computer Vision (WACV)}, January 2021, pp. 576--585.

\bibitem{gonzalez2020comparing}
S.~Gonz{\'a}lez-Carvajal and E.~C. Garrido-Merch{\'a}n, ``Comparing bert
  against traditional machine learning text classification,'' \emph{arXiv
  preprint arXiv:2005.13012}, 2020.

\bibitem{atkinson_1985}
J.~M. Atkinson, \emph{Public speaking and audience responses: some techniques
  for inviting applause}, ser. Studies in Emotion and Social Interaction.\hskip
  1em plus 0.5em minus 0.4em\relax Cambridge University Press, 1985, p.
  370–410.

\bibitem{heritage1986generating}
J.~Heritage and D.~Greatbatch, ``Generating applause: A study of rhetoric and
  response at party political conferences,'' \emph{American journal of
  sociology}, vol.~92, no.~1, pp. 110--157, 1986.

\bibitem{bull2011invitations}
P.~Bull and O.~Feldman, ``Invitations to affiliative audience responses in
  japanese political speeches,'' \emph{Journal of Language and Social
  Psychology}, vol.~30, no.~2, pp. 158--176, 2011.

\bibitem{liu2017fostering}
Z.~Liu, A.~Xu, M.~Zhang, J.~Mahmud, and V.~Sinha, ``Fostering user engagement:
  Rhetorical devices for applause generation learnt from ted talks,'' in
  \emph{Proceedings of the International AAAI Conference on Web and Social
  Media}, vol.~11, no.~1, 2017.

\bibitem{bertero2016deep}
D.~Bertero and P.~Fung, ``Deep learning of audio and language features for
  humor prediction,'' in \emph{Proceedings of the Tenth International
  Conference on Language Resources and Evaluation (LREC'16)}, 2016, pp.
  496--501.

\bibitem{bertero2016multimodal}
------, ``Multimodal deep neural nets for detecting humor in tv sitcoms,'' in
  \emph{2016 IEEE Spoken Language Technology Workshop (SLT)}.\hskip 1em plus
  0.5em minus 0.4em\relax IEEE, 2016, pp. 383--390.

\bibitem{kayatani2021laughing}
Y.~Kayatani, Z.~Yang, M.~Otani, N.~Garcia, C.~Chu, Y.~Nakashima, and
  H.~Takemura, ``The laughing machine: Predicting humor in video,'' in
  \emph{Proceedings of the IEEE/CVF Winter Conference on Applications of
  Computer Vision}, 2021, pp. 2073--2082.

\bibitem{choube2020punchline}
S.~Choube, Akshat and Mohammad, ``Punchline detection using context-aware
  hierarchical multimodal fusion,'' in \emph{Proceedings of the 2020
  International Conference on Multimodal Interaction}, 2020, pp. 675--679.

\bibitem{hasan2021humor}
M.~K. Hasan, S.~Lee, W.~Rahman, A.~Zadeh, R.~Mihalcea, L.-P. Morency, and
  E.~Hoque, ``Humor knowledge enriched transformer for understanding multimodal
  humor,'' in \emph{Proceedings of the AAAI Conference on Artificial
  Intelligence}, vol.~35, no.~14, 2021, pp. 12\,972--12\,980.

\bibitem{fan2020humor}
X.~Fan, H.~Lin, L.~Yang, Y.~Diao, C.~Shen, Y.~Chu, and Y.~Zou, ``Humor
  detection via an internal and external neural network,''
  \emph{Neurocomputing}, vol. 394, pp. 105--111, 2020.

\bibitem{fan2020phonetics}
X.~Fan, H.~Lin, L.~Yang, Y.~Diao, C.~Shen, Y.~Chu, and T.~Zhang, ``Phonetics
  and ambiguity comprehension gated attention network for humor recognition,''
  \emph{Complexity}, vol. 2020, 2020.

\bibitem{navarretta2017prediction}
C.~Navarretta, ``Prediction of audience response from spoken sequences, speech
  pauses and co-speech gestures in humorous discourse by {Barack Obama},'' in
  \emph{2017 8th IEEE International Conference on Cognitive Infocommunications
  (CogInfoCom)}.\hskip 1em plus 0.5em minus 0.4em\relax IEEE, 2017, pp.
  000\,327--000\,332.

\bibitem{ruf2020creating}
V.~Ruf and C.~Navarretta, ``Creating a corpus of gestures and predicting the
  audience response based on gestures in speeches of donald trump,'' in
  \emph{Proceedings of the 12th Language Resources and Evaluation Conference},
  2020, pp. 1081--1088.

\bibitem{devlin2018bert}
J.~Devlin, M.-W. Chang, K.~Lee, and K.~Toutanova, ``{BERT}: Pre-training of
  deep bidirectional transformers for language understanding,'' \emph{arXiv
  preprint arXiv:1810.04805}, 2018.

\bibitem{yoon2019robots}
Y.~Yoon, W.-R. Ko, M.~Jang, J.~Lee, J.~Kim, and G.~Lee, ``Robots learn social
  skills: End-to-end learning of co-speech gesture generation for humanoid
  robots,'' in \emph{2019 International Conference on Robotics and Automation
  (ICRA)}.\hskip 1em plus 0.5em minus 0.4em\relax IEEE, 2019, pp. 4303--4309.

\end{thebibliography}

\end{document}